\documentclass[pra,aps,twocolumn,showpacs,superscriptaddress,10pt]{revtex4-1}

\usepackage{bm}
\usepackage{amsmath,amssymb}
\usepackage{graphicx}
\usepackage{color}
\usepackage{hyperref}
\usepackage[all]{hypcap} 
\usepackage{dcolumn}

\newcommand{\fak}[0]{1.}  

\newcommand{\mytitle}{Driving Rabi oscillations at the giant dipole resonance in xenon}

\definecolor{MyDarkGreen}{rgb}{0,0.6,0}
\definecolor{MyDarkBlue}{rgb}{0,0,0.8}
\definecolor{MyDarkRed}{rgb}{0.6,0,0.3}
\hypersetup{breaklinks=true, colorlinks=true,plainpages=true, linktocpage=true, linkcolor=MyDarkBlue, citecolor=MyDarkGreen, urlcolor=MyDarkRed, pdfborder={0 0 0},%
pdfauthor={Stefan Pabst},%
pdfsubject={Research article},
pdftitle={\mytitle}%
}

\newcommand{\ket}[1]{\left|#1\right>}

\newcommand{\sket}[1]{\left|#1\right)}
\newcommand{\sbra}[1]{\left(#1\right|}

\begin{document}

\title{\mytitle} 

\author{Stefan Pabst}
\email[]{stefan.pabst@cfa.harvard.edu}
\affiliation{Center for Free-Electron Laser Science, DESY, Notkestrasse 85, 22607 Hamburg, Germany}
\affiliation{ITAMP, Harvard-Smithsonian Center for Astrophysics, 60 Garden Street, Cambridge, MA 02138, USA}

\author{Daochen Wang}
\affiliation{Center for Free-Electron Laser Science, DESY, Notkestrasse 85, 22607 Hamburg, Germany}
\affiliation{DAMTP, University of Cambridge, Wilberforce Road, Cambridge CB3 0WA, United Kingdom}

\author{Robin Santra}
\email[]{robin.santra@cfel.de}
\affiliation{Center for Free-Electron Laser Science, DESY, Notkestrasse 85, 22607 Hamburg, Germany}
\affiliation{Department of Physics, University of Hamburg, Jungiusstrasse 9, 20355 Hamburg, Germany}

\date{\today}


\begin{abstract}
Free-electron lasers (FELs) produce short and very intense light pulses in the XUV and x-ray regimes.
We investigate the possibility to drive Rabi oscillations in xenon with an intense FEL pulse by using the unusually large dipole strength of the giant-dipole resonance (GDR).
The GDR decays within less than 30~as due to its position, which is above the $4d$ ionization threshold.
We find that intensities around 10$^{18}$~W/cm$^2$ are required to induce Rabi oscillations with a period comparable to the lifetime.
The pulse duration should not exceed 100~as because xenon will be fully ionized within a few lifetimes. 
Rabi oscillations reveal themselves also in the photoelectron spectrum in form of Autler-Townes splittings extending over several tens of electronvolt.
\end{abstract} 

\maketitle


\section{Introduction}

In the last decade, with the emergence of free-electron lasers (FELs)~\cite{AcAs-NatPhot-2007,EmAk-nphoto-2010,IsAo-nphoto-2012,MuNi-NatPhoton-2012}, the door has been opened for studying multiphoton processes in the XUV and x-ray regimes.
Complex ionization dynamics, resulting from a non-trivial interplay of photoabsorption and inner-shell processes such as Auger decay, have been experimentally found when exposing atoms~\cite{YoKa-Nature-2010,DoRo-PRL-2011,RoRy-Nature-2012}, molecules~\cite{HoFa-PRL-2010,CrGl-PRL-2010}, and clusters~\cite{BoTh-NJP-2010,ThHe-PRL-2012} to these high-fluence and high-intensity FEL pulses. 

From the extensive works in the last 40 years~\cite{Scully-book}, it is well known that intense optical pulses with frequencies resonant with internal transitions cause the system to behave in completely new ways, resulting in novel effects such as Rabi oscillations, electromagnetically-induced transparency (EIT), lasing without inversion, and population trapping.
These processes have been primarily studied in the outer-valence shells of atomic systems, which are accessible with optical frequencies and the physical realization is quite close to ideal isolated two-, three-, and multi-level systems~\cite{WuEz-PRL-1977,SeWi-PRA-1996}.
With the arrival of FELs, these processes have been extended to the XUV and x-ray regimes: namely stimulated emission~\cite{RoRy-Nature-2012}, Rabi oscillations~\cite{KaKr-PRL-2011,RoSa-PRA-2012}, and processes similar to EIT~\cite{BuSa-PRL98-2007,GlHe-NatPhys-2010}.

At XUV and x-ray photon energies ($\omega > 10$~eV), atomic bound-bound transitions must involve inner-valence and core shells.
Once an electron is removed from a core orbital, the system wants to 'relax' by filling this hole via spontaneous emission (more likely for heavy atoms) or Auger decay (more likely for light atoms)~\cite{XDB}.
In order to compete with the relaxation processes of the system, the time scale of the light-driven processes has to be comparable or ideally faster than the time scale of the relaxation.
If this is the case, the light-driven process will not just dominate over the relaxation processes but it also significantly alters the relaxation processes themselves. 
The influence of light-driven processes on x-ray fluorescence~\cite{CaKe-PRA-2012} and Auger decay~\cite{RoSa-PRA-2008,RoSa-PRA-2012,CaGe-PRA-2010,NiCo-PRA-2011,SaAd-PRA-2011,HaMc-PRA-2014,DeCh-PRA-2011,*MuDe-JPB-2015,ChNa-PRA-2015} has been theoretically studied in recent years. 

In this work, we investigate the possibility to induce Rabi oscillations involving the giant-dipole resonance (GDR) in xenon. 
The GDR is located at around 100~eV above the ground state and, therefore, also above the $4d$ ionization threshold.
Even though the electron is ultimately ionized at this energy, it is bound for a very short time in the vicinity of the atom leading to an enhanced dipole transition strength which gave the GDR its name~\cite{FaCo-RMP-1968,*St-Springer-1980,KrPa-AJP-2014}.
The unusually large dipole strength is beneficial in two ways.
First, it induces fast Rabi oscillations that can compete with the short lifetimes ($<30$~as) of the GDR states.
Second, it ensures that other ionization pathways (out of other sub-shells), which do not involve the GDR, are much weaker and are not of high relevance.
In high-harmonic generation, this large dipole moment can be used to significantly boost the high-harmonic yield around 100~eV~\cite{FrSt-PRL-2009,ShVi-NatPhys-2011,PaSa-PRL-2013}.

Theoretical studies~\cite{We-JPB-1973,ChPa-PRA-2015} have found that the GDR consists of two sub-resonances.
Recently, an experiment~\cite{MaKa-NatComm-2015} has seen first indications of this sub-structure in the XUV two-photon above-threshold-ionization (ATI) spectrum of xenon.
Another goal of this study is, therefore, to investigate whether Rabi oscillations can uncover this sub-structure as well.

In the following, we present in Sec.~\ref{sec:theory} our theoretical model. 
In Sec.~\ref{sec:results} we estimate which pulses are needed to induce Rabi oscillations, study the population dynamics, and investigate how Rabi oscillations affect the $4d$ hole population and the photoelectron spectrum.
If not noted otherwise, atomic units are used throughout the paper.

\section{Theory}
\label{sec:theory}


We use our time-dependent configuration interaction singles (TDCIS) approach~\cite{GrSa-PRA-2010,*Pa-EPJST-2013}, which we have successfully applied in the strong-field~\cite{PaGr-PRA-2012,*PaSa-JPB-2014,PaSa-PRL-2013,WiGo-Science-2011}, XUV~\cite{PaSa-PRL-2011,KrPa-AJP-2014,KaPa-PRA-2014,ChPa-PRA-2015,MaKa-NatComm-2015}, and x-ray regimes~\cite{SyPa-PRA-2012,TiKa-JPB-2015}.
The $N$-body wavefunction ansatz for TDCIS reads
\begin{align}
 \label{eq:tdcis}
 \ket{\Psi(t)}
 &=
 \alpha_0(t) \sket{\Phi_0}
 +
 \sum_{ai}  \alpha^a_i(t) \sket{\Phi^a_i}
 ,
\end{align}
where $\Phi_0$ is the Hartree-Fock (HF) ground state, and $\Phi^a_i$ is a one-particle-one-hole excitation where one electron is excited from orbital $i$ into orbital $a$.
The Hamiltonian is the exact non-relativistic $N$-body Hamiltonian,
\begin{align}
  \label{eq:ham}
  \hat H(t)
  &=
  \hat H_0 + \hat H_1 + A(t)\,\hat p - E_\text{HF}
  ,
\end{align}
which is partitioned into four parts: (i) the Fock operator, $\hat H_0$, describing non-interacting electrons in the HF mean-field potential plus a complex-absorbing potential, (ii) the residual Coulomb interaction, $\hat H_1$, capturing the electron-electron interactions that go beyond the HF mean-field picture, (iii) the light-matter interaction, $A(t)\,\hat p$, in the velocity form of the dipole approximation with $A(t)$ being the vector potential and $\hat p$ being the momentum operator, and (iv) the Hartree-Fock energy, $E_\text{HF}$, which shifts the spectrum such that the HF ground state has the energy $0$.

The residual Coulomb interactions, $\hat H_1$, can be grouped into two classes: intrachannel and interchannel coupling.
Intrachannel coupling, $\sbra{\Phi^a_i} \hat H_1 \sket{\Phi^b_i}$, corrects the mean-field potential due to the missing electron in the atom, and leads to a long-range, $-1/r$, potential for the photoelectron.
Interchannel coupling, $\sbra{\Phi^a_i} \hat H_1 \sket{\Phi^b_j}$ ($i\neq j$), describes the interaction where the excited electron changes the ionic states $i$.
This interaction leads correlated electron dynamics that can significantly change the overall response of the system~\cite{PaSa-PRL-2013} and has large effects on coherence properties~\cite{PaSa-PRL-2011}. 

Combining Eqs.~\eqref{eq:tdcis} and \eqref{eq:ham}, we find the equation of motion for the time-dependent CIS coefficients,
\begin{subequations}
\label{eq:eom}
\begin{align}
  \label{eq:eom.1}
  i\partial_t \, \alpha_0(t)
  &=
  A(t)\, \sum_{a,i} \sbra{\Phi_0} \hat p \sket{\Phi^a_i}
  \alpha^a_i(t)
  ,
  \\\nonumber
  \label{eq:eom.2}
  i\partial_t \, \alpha^a_i(t)
  &= 
  (\varepsilon_a-\varepsilon_i) \, \alpha^a_i(t)
  +
  \sum_{b,j}
    \sbra{\Phi^a_i} \hat H_1 \sket{\Phi^b_j}
    \alpha^b_j(t)
  \\ &
  +
  A(t)\, \Big(
    \alpha_0(t)
    \sbra{\Phi^a_i} \hat p \sket{\Phi_0}
    \! + \!
    \sum_{jb} \sbra{\Phi^a_i} \hat p \sket{\Phi^b_j}
    \alpha^b_j(t)
  \Big)
  ,
\end{align}
\end{subequations}
where $\sket{\Phi^a_i}$ and $\sbra{\Phi^a_i}$ are right and left eigenstates of the non-hermitian Fock operator $\hat H_0$, respectively (see Ref.~\cite{GrSa-PRA-2010}).
The energies of the occupied and virtual orbitals are given by $\varepsilon_i$ and $\varepsilon_a$, respectively.

%
%
 The existence of the GDR in xenon can be understood in a single-particle picture with a central model potential~\cite{St-PRA-1970,St-Springer-1980}.
 CIS intrachannel already improves the description of the GDR in terms of energy position and spectral width~\cite{St-PRA-1970,Pa-EPJST-2013}.
 However, many-body effects have to be taken into account in order to reproduce the correct position and width of the GDR. 
 Interchannel interactions, $\sbra{\Phi^a_i} \hat H_1 \sket{\Phi^b_j}$, within CIS improve greatly the description of the GDR in comparison to intrachannel CIS~\cite{St-PRA-1970,Pa-EPJST-2013,KrPa-AJP-2014}.
 To obtain an even better description, electronic correlations of higher order, mostly double excitations, are needed~\cite{St-Springer-1980}.
 
 Recently, it has been shown that interchannel interactions lead to the emergence of a second dipole-allowed resonance state within the GDR~\cite{ChPa-PRA-2015}.
 This second resonance is centered at 112~eV and lives only for 11~as ($\Gamma=58.2$~eV).
 It is primarily responsible for the spectrally broad GDR feature in the photoionization cross section.
 In comparison, the first GDR resonance, which emerges already from a one-particle picture due to a shape resonance, is centered at 73~eV and has a lifetime of 27~as ($\Gamma=24.7$~eV).

\section{Results}
\label{sec:results}

We use the {\sc xcid} program~\footnote{S. Pabst, L. Greenman, A. Karamatskou, Y.-J. Chen, A. Sytcheva, O. Geffert, R. Santra--\textsc{xcid} program package for multichannel ionization dynamics, DESY, Hamburg, Germany, 2015, Rev. 1790} (with the following numerical parameters in~\footnote{
A pseudo-spectral grid with a radial box size of 100~$a_0$, 500 grid points, and a mapping parameter of $\zeta=0.5$ are used. 
The complex absorbing potential starts at 70~$a_0$ and has a strength of $\eta=0.002$.
The maximum angular momentum is 6 and Hartree-Fock orbitals up to an energy of 200~$E_h$ are considered. 
The propagation method is Runge-Kutta 4 with a time step $dt=0.005$~a.u.
All $4d, 5s$, and $5p$ orbitals are active in the calculations.
})
to verify that we can induce Rabi oscillations using the GDR resonances.
With TDCIS, we do not simplify the problem to a two-level system, and we explicitly include all other ionization mechanisms (that lead to singly ionized xenon).
Furthermore, the GDR is properly described as a result of discrete states in the continuum~\cite{ChPa-PRA-2015} which can be modified by the intense XUV pulse itself.
This may lead to trends that deviate from the weak-field behavior.

%
%

\subsection{Population dynamics}
\label{sec:results.pop-dyn}

\begin{figure}[t!]
  \label{fig:rabi.time}
  \includegraphics[clip,width=\fak\linewidth]{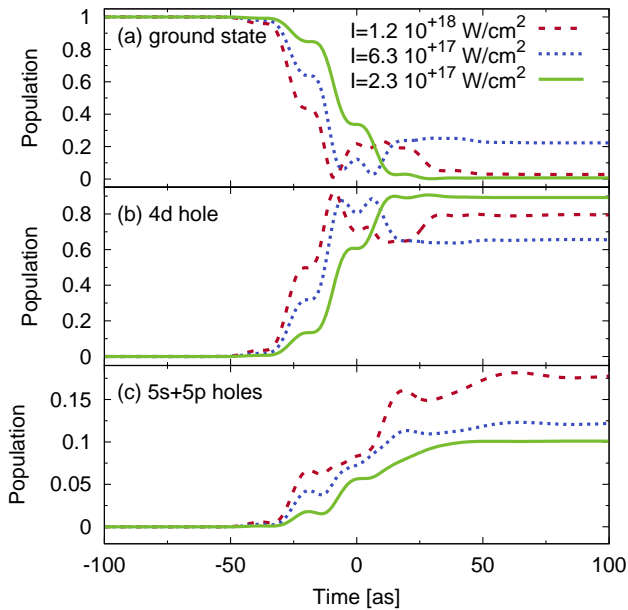} 
  \caption{(color online) The population of (a) the ground state, (b) a hole in the $4d$ shell, and (c) a hole in the $5s$ or $5p$ shell in atomic xenon.
  The instantaneous peak intensity of the 36~as long (FWHM) Gaussian pulse with a center photon energy of 109~eV is varied from $2.3\cdot 10^{17}$~W/cm$^2$ to $1.2\cdot 10^{18}$~W/cm$^2$.}
\end{figure}

First, we have a look at the dynamics of the ground state population (cf. Fig.~\ref{fig:rabi.time}a) and of the hole populations of different sub-shells of xenon (cf. Fig.~\ref{fig:rabi.time}b-c) as it is exposed to a pulse with a center photon energy of 109~eV ($=4$~a.u.) and a FWHM duration (with respect to the intensity profile) of 36~as ($=1.5$~a.u.).
With intensities above $10^{17}$~W/cm$^2$, we basically fully ionize xenon.
As we increase the intensity, the ground state population and the individual hole populations start to show oscillatory behavior and do not monotonically increase as in the ``low'' intensity limit (cf. $2.3\cdot10^{17}$~W/cm$^2$ results in Fig.~\ref{fig:rabi.time}).
These are clear indications that we successfully induce Rabi oscillations.
Rabi oscillations are strongly damped due to the short lifetime of the GDR states leading to a large amount of irreversible ionization per Rabi cycle.
Note that for relatively low intensities the hole populations grow monotonically with time. 

In Fig.~\ref{fig:rabi.time}, we see that mainly the $4d$ shell is depopulated as expected from the character of the GDR, which mainly corresponds to a configuration where a $4d$ electron is excited into the $l=3$ continuum.
The ionization of the $5s$ and $5p$ shells is ten times smaller but with 10\% ionization probability it is large enough that it should be taken into account.


\subsection{Final hole population}
\label{sec:results.pop-final}

Unfortunately, it is very hard to monitor the $4d$ hole population as a function of time in an experiment.
With attosecond transient absorption spectroscopy~\cite{WiGo-Science-2011,PaSy-PRA-2012}, it is in principle possible to do so.
The pump-probe delay between the pulses can already be controlled on a few attosecond scale, but the challenge lies in the generation of a probe pulse with a duration of a few attoseconds to achieve the required time resolution. 
This comes on top of the already challenging requirements for the XUV pump pulse.
It is, therefore, easier to probe Rabi oscillations by measuring the final $4d$ hole population or the photoelectron spectrum (see Sec.~\ref{sec:results.esepc}).

\begin{figure}[t!]
  \label{fig:finalpop.varyE}
  \includegraphics[clip,width=\fak\linewidth]{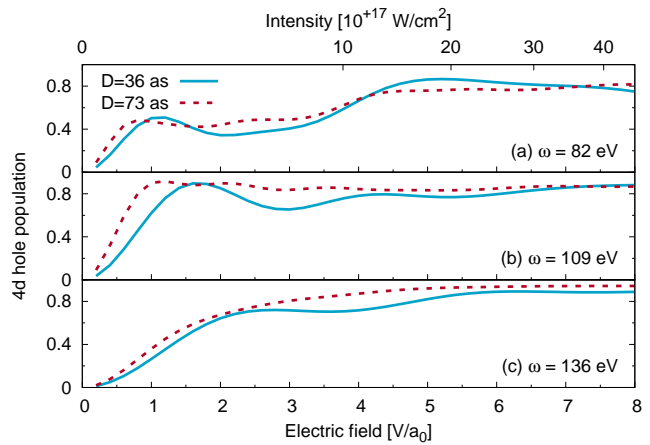} 
  \caption{(color online) The final $4d$ hole population as a function of pulse intensity for different center photon energies $\omega$ (a-c), and two pulse durations, 36~as (blue solid) and 73~as (red dashed).
  } 
\end{figure}

The $4d$ hole itself is not stable and will decay mainly via Auger decay.
The lifetime of the $4d$ hole is around 6~fs. 
Therefore, we can safely neglect the hole decay during the few attoseconds the XUV pulse drives Rabi oscillations. 
The hole will, however, eventually decay. 
The Auger electron yield is, therefore, directly related to the final $4d$ hole population after the XUV pulse. 
Also the Auger electron spectrum should not be affected by the XUV pulse as the $4d$ hole decays predominantly after the pulse under field-free conditions. 

In Fig.~\ref{fig:finalpop.varyE}, the final $4d$ hole population (after the pulse) is shown as a function of the pulse intensity for the XUV photon energies, (a) $\omega=82$~eV [$=3$~a.u.], (b) $\omega=109$~eV [$=4$~a.u.], and (c) $\omega=136$~eV [$=5$~a.u.], as well as for the pulse durations, (solid blue line) 36~as [$=1.5$~a.u.] and (dashed red line) 73~as [$=3.0$~a.u.].
We see, especially for the 36~as pulse, that the population has oscillatory behavior and does not monotonically grow with the pulse intensity---a clear indication of Rabi oscillations.
The more rapid modulations in the final population for longer pulses is also consistent with Rabi oscillations.
The probability to be in an excited state after the pulse is given by $\sin^2(\Omega_\text{eff}T)$, where $\Omega_\text{eff}=T^{-1}\int_{-\infty}^\infty\!\! dt\ d\,{\cal E}(t)=d\,{\cal E}_\text{max} \sqrt{\pi/(2\ln(2))}$ is the pulse-averaged Rabi frequency, ${\cal E}(t)$ is the pulse envelope, $d$ is the dipole transition strength, and $T$ is the duration of the pulse.

Even though the variations are faster for longer pulses, the visibility decreases with pulse length.
Due to the short lifetimes of the GDR states, which are below 30~as, almost all electrons are irreversibly ionized after 30~as, and the hole population is always close to unity whether or not Rabi oscillations were induced. 

The hole creation is most dominant at $\omega=109$~eV where the photoionization cross-section is largest (for the three photon energies shown).
At $\omega=136$~eV, the hole creation increases almost monotonically with the electric field.
Almost no oscillatory behavior is visible indicating that at this photon energy we do not hit a resonance and ionization is predominantly irreversible.
At $\omega=82$~eV, the first peak in the hole population appears quite early followed by a relatively long plateau extending up to $10^{17}$~W/cm$^2$.
This trend is not fully consistent with the Rabi-oscillation picture of a two-level system.
But xenon at these XUV photon energies cannot be described as a two-level system.

\begin{figure}[t!]
  \label{fig:finalpop.varyE-w}
  \includegraphics[clip,width=\fak\linewidth]{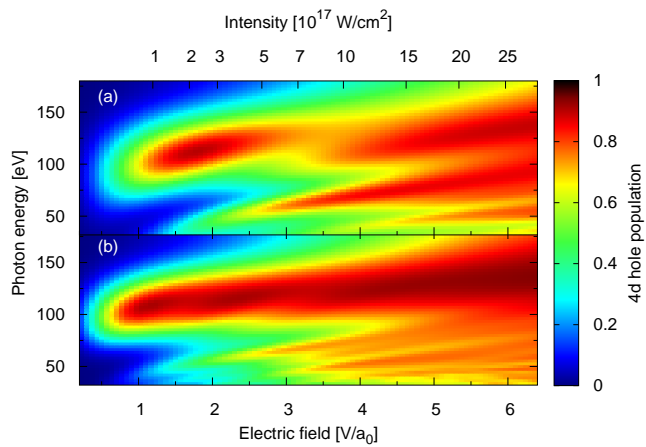} 
  \caption{(color online) The final $4d$ hole population as a function of pulse intensity and center pulse photon energy.
  The pulse duration is (a) 36~as and (b) 73~as.
  } 
\end{figure}

As we see in Fig.~\ref{fig:finalpop.varyE-w}, where the final $4d$ hole population is shown as a function of intensity and driving photon energy, the photon energy where the $4d$ shell is most strongly ionized shifts to higher energies as the intensity increases.
This explains why for $\omega=82$~eV in  Fig.~\ref{fig:finalpop.varyE}a) a plateau appears. 
At low intensities, it probes the lower end of the GDR.
At higher intensities, the GDR moves to higher energies and the dipole transition strength drops, counter-balancing the increase in field strength.

The energy shift of the GDR is a result of the very intense XUV pulse, which dresses the excited and the continuum states.
The polarizability of a flat continuum is given by the ponderomotive potential $U_p=\frac{E^2}{4\omega^2}$ and shifts the continuum to higher energies~\cite{DeKr-book}.
At $E=2$~a.u. and $\omega=109$~eV, this yields an energy shift of less than 2~eV.
The observed energy shift of around 20~eV is much larger.
Furthermore, the energy shifts depend rather linearly than quadratically on $E$.
As we will see in Sec.~\ref{sec:results.esepc}, the kinetic energy of the photoelectron even increases or decreases with intensity depending on whether the photon energy is above or below the GDR, respectively. 
This clearly shows that the continuum in xenon is structured around the GDR and is not flat.
Furthermore, when $z\,E\gg\omega$ linear energy shifts are expected~\cite{DeKr-book}.
In Fig.~\ref{fig:finalpop.varyE-w}, we find the linear energy shifts especially at intensities above 10$^{18}$~W/cm$^2$ where $z\,E>\omega$.


For the short 36~as pulse, the effect of Rabi oscillations is nicely visible. 
At $\omega\sim 100$--150~eV, the $4d$ hole population shows a local minimum around $7-11\cdot 10^{17}$~W/cm$^2$ and an island of enhanced ionization emerges at lower intensities.
For the longer pulse, more Rabi oscillation occur as the intensity increases and, therefore, two islands of enhanced ionization occur. 
Again, the visibility is greatly reduced for longer pulses as explained above.

At photon energies below 100~eV, no signs of Rabi oscillations can be seen. 
Even though the dipole strength falls off quite symmetrically around 100~eV in the weak-field regime, the dressing of the continuum states creates a clear preference for higher energies as the field strength increases.
Furthermore, we see no indication of the two GDR subresonances.
Rabi oscillations seem to be only sensitive to the overall dipole strength.
Note that at each photon energies the ground state is coupled resonantly to an excited/continuum state in contrast to a two-level system where only one excited state at a specific energy exists. 


Another general trend that we can observe in Fig.~\ref{fig:finalpop.varyE-w} is the broadening of the spectral feature with intensity.
This indicates, on the one hand, that the lifetimes of the excited states are decreasing.
On the other hand, the spectral broadening is also a direct consequence of saturation of ionization.
We already saw in Figs.~\ref{fig:rabi.time} and \ref{fig:finalpop.varyE} that at these intensities we basically fully ionize the system.

Also new multiphoton ionization pathways start to emerge for photon energies below the $4d$ ionization threshold and at intensities above $2\cdot10^{17}$~W/cm$^2$ ($E\geq 1.5$~a.u.).
The ionization is enhanced for specific combinations of field strengths and photon energies.
The photon energy for the ionization enhancement increases as the field strength increases indicating field-dressing effects are involved in these new pathways. 
The origin might be, therefore, quite similar to Freeman resonances known from ATI~\cite{FrBu-PRL-1987}, where intermediate states are shifted into resonance by the intense pulse.

\subsection{Photoelectron spectrum}
\label{sec:results.esepc}

Rabi oscillations should be also visible in the photoelectron spectrum.
In a time domain picture, the field-driven coupling between two states leads to Rabi oscillations.
In an energy domain picture, the excited state (also the ground state) splits into two states, known as the Autler-Townes doublet, which are separated by the Rabi frequency $\Omega$~\cite{FlIm-RMP-2005}. 
In our case, the excited state is in the continuum, and the energy splitting should be directly imprinted in the kinetic energy of the photoelectron.

\begin{figure}[t!]
  \label{fig:espec.f4}
  \includegraphics[clip,angle=270,width=\fak\linewidth]{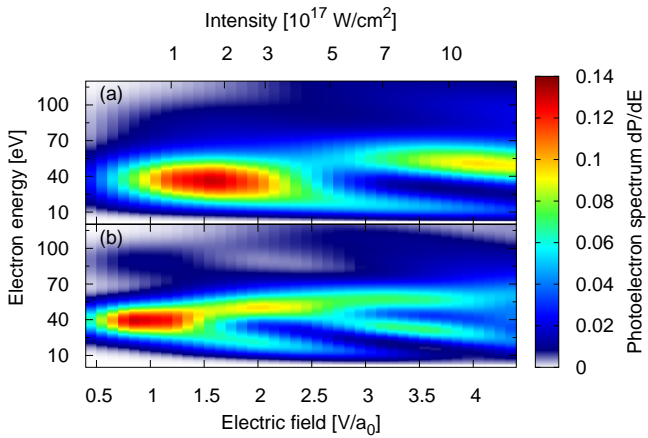} 
  \caption{(color online) The photoelectron spectrum as a function of the pulse intensity.
  The photon energy is centered at 109~eV and the pulse duration is (a) 36~as and (b) 73~as.
  } 
\end{figure}

In Fig.~\ref{fig:espec.f4}, the total photoelectron spectrum is shown as a function of intensity for (a) a 36~as and (b) a 73~as pulse with $\omega=109$~eV.
We clearly see the Autler-Townes doublet for both pulse durations.
The lower energy branch of the Autler-Townes doublet approaches 0~eV and does not survive at high intensities as the kinetic energy of the electron has to be positive. 
At higher intensities, the ionization dynamics become less periodic and additional peaks in the photoelectron spectrum occur~\cite{RzZa-PRA-1985,RoSa-PRA-2008}.


\begin{figure}[t!]
  \label{fig:espec.f3-f5}
  \includegraphics[clip,angle=270,width=\fak\linewidth]{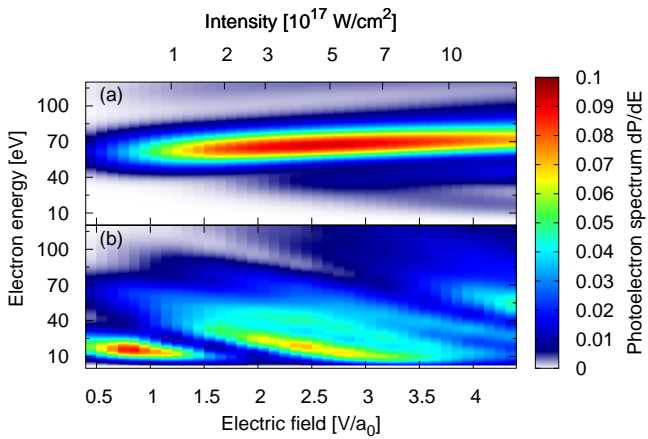} 
  \caption{(color online) The photoelectron spectrum as a function of the pulse intensity.
  The pulse has a duration of 73~as and the photon energy is centered (a) at 136~eV and (b) at 82~eV.
  }
\end{figure}

Once the driving photon energy moves away from $110$~eV, no indication of Rabi oscillations are seen in the $4d$ hole population (see Fig.~\ref{fig:finalpop.varyE-w}).
The same is true for the photoelectron spectrum.
In Fig.~\ref{fig:espec.f3-f5}, the photoelectron spectrum is shown for (a) $\omega=136$~eV and (b) $\omega=82$~eV as the pulse intensity is varied.
For $\omega=136$~eV, the kinetic energy of the electron slightly increases with intensity.
The coupling to the lower lying GDR resonance pushes the continuum states above the GDR to lower energies. 
At $\omega=163$~eV ($=6$~a.u.)---these results are not shown---the energy separation to the GDR is large and the coupling can be neglected and the kinetic energy of the photoelectron does not significantly change with intensity.

For photon energies below the GDR (see Fig.~\ref{fig:espec.f3-f5}b), the trend is reversed and the kinetic energy of the electron decreases with intensity.
This clearly shows that the polarization of the continuum at these frequencies is strongly affected by the GDR and cannot be considered flat.
For a flat continuum, the kinetic energy of the photoelectron spectrum would always decrease by the ponderomotive potential ($\propto\omega^{-2}$) and would never increase as $\omega$ increase.

\section{conclusion}
\label{sec:conclusion}
We have investigated to which extent intense FEL pulses could be used to drive Rabi oscillations between the neutral ground state and the GDR in xenon.
We found that intensities around $10^{18}$~W/cm$^2$ are needed to see an impact of Rabi oscillations on the $4d$ hole population.
We could find indications that Rabi oscillation can be used to uncover the substructure of the GDR, i.e., the two dipole-allowed resonances at 73~eV and 112~eV~\cite{ChPa-PRA-2015}.
There are two ways how Rabi oscillations can be observed: (1) by measuring the $4d$ hole population via the Auger electron that gets emitted when the $4d$ hole decays, or (2) via the photoelectron spectrum in the form of Autler-Townes splitting.

To see Rabi oscillations in the $4d$ hole population, a short pulse duration comparable to the lifetimes of the GDR states is needed.
For the photoelectron spectrum, a longer pulse seems to be beneficial as the Autler-Townes splitting is spectrally better visible. 
The energy shifts of the photoelectron show clearly that the continuum around the GDR is structured and highly polarizable and cannot be assumed flat. 
With seeded FELs, pulse intensities and pulse durations are in reach to induce Rabi oscillations that are driven by XUV light.

\acknowledgments
S.P. is funded by the Alexander von Humboldt Foundation and by the NSF through a grant to ITAMP.

\bibliographystyle{apsrev4-1-etal}
\bibliography{amo,books}

\begin{thebibliography}{54}%
\makeatletter
\providecommand \@ifxundefined [1]{%
 \@ifx{#1\undefined}
}%
\providecommand \@ifnum [1]{%
 \ifnum #1\expandafter \@firstoftwo
 \else \expandafter \@secondoftwo
 \fi
}%
\providecommand \@ifx [1]{%
 \ifx #1\expandafter \@firstoftwo
 \else \expandafter \@secondoftwo
 \fi
}%
\providecommand \natexlab [1]{#1}%
\providecommand \enquote  [1]{``#1''}%
\providecommand \bibnamefont  [1]{#1}%
\providecommand \bibfnamefont [1]{#1}%
\providecommand \citenamefont [1]{#1}%
\providecommand \href@noop [0]{\@secondoftwo}%
\providecommand \href[0]{\begingroup \@sanitize@url \@href}%
\providecommand \@href[1]{\@@startlink{#1}\@@href}%
\providecommand \@@href[1]{\endgroup#1\@@endlink}%
\providecommand \@sanitize@url [0]{\catcode `\\12\catcode `\$12\catcode
  `\&12\catcode `\#12\catcode `\^12\catcode `\_12\catcode `\%12\relax}%
\providecommand \@@startlink[1]{}%
\providecommand \@@endlink[0]{}%
\providecommand \url  [0]{\begingroup\@sanitize@url \@url }%
\providecommand \@url [1]{\endgroup\@href {#1}{\urlprefix }}%
\providecommand \urlprefix  [0]{URL }%
\providecommand \Eprint [0]{\href }%
\providecommand \doibase [0]{http://dx.doi.org/}%
\providecommand \selectlanguage [0]{\@gobble}%
\providecommand \bibinfo  [0]{\@secondoftwo}%
\providecommand \bibfield  [0]{\@secondoftwo}%
\providecommand \translation [1]{[#1]}%
\providecommand \BibitemOpen [0]{}%
\providecommand \bibitemStop [0]{}%
\providecommand \bibitemNoStop [0]{.\EOS\space}%
\providecommand \EOS [0]{\spacefactor3000\relax}%
\providecommand \BibitemShut  [1]{\csname bibitem#1\endcsname}%
\let\auto@bib@innerbib\@empty
\bibitem [{\citenamefont {Ackermann}\ \emph {et~al.}(2007)\citenamefont
  {Ackermann}, \citenamefont {Asova}, \citenamefont {Ayvazyan}, \citenamefont
  {Azima}, \citenamefont {Baboi}, \citenamefont {B\"ahr}, \citenamefont
  {Balandin}, \citenamefont {Beutner}, \citenamefont {Brandt}, \citenamefont
  {Bolzmann}, \citenamefont {Brinkmann}, \citenamefont {Brovko} \emph
  {et~al.}}]{AcAs-NatPhot-2007}%
  \BibitemOpen
  \bibfield  {author} {\bibinfo {author} {\bibfnamefont {W.}~\bibnamefont
  {Ackermann}}, \bibinfo {author} {\bibfnamefont {G.}~\bibnamefont {Asova}},
  \bibinfo {author} {\bibfnamefont {V.}~\bibnamefont {Ayvazyan}},  \emph
  {et~al.},\ }\href{\doibase 10.1038/nphoton.2007.76} {\bibfield  {journal}
  {\bibinfo  {journal} {Nat. Phot.}\ }\textbf {\bibinfo {volume} {1}},\
  \bibinfo {pages} {336} (\bibinfo {year} {2007})}\BibitemShut {NoStop}%
\bibitem [{\citenamefont {Emma}\ \emph {et~al.}(2010)\citenamefont {Emma},
  \citenamefont {Akre}, \citenamefont {Arthur}, \citenamefont {Bionta},
  \citenamefont {Bostedt}, \citenamefont {Bozek}, \citenamefont {Brachmann},
  \citenamefont {Bucksbaum}, \citenamefont {Coffee}, \citenamefont {Decker},
  \citenamefont {Ding}, \citenamefont {Dowell}, \citenamefont {Edstrom},
  \citenamefont {Fisher}, \citenamefont {Frisch}, \citenamefont {Gilevich},
  \citenamefont {Hastings}, \citenamefont {Hays}, \citenamefont {Hering},
  \citenamefont {Huang}, \citenamefont {Iverson}, \citenamefont {Loos},
  \citenamefont {Messerschmidt}, \citenamefont {Miahnahri}, \citenamefont
  {Moeller}, \citenamefont {Nuhn}, \citenamefont {Pile}, \citenamefont
  {Ratner}, \citenamefont {Rzepiela}, \citenamefont {Schultz}, \citenamefont
  {Smith}, \citenamefont {Stefan}, \citenamefont {Tompkins}, \citenamefont
  {Turner}, \citenamefont {Welch}, \citenamefont {White}, \citenamefont {Wu},
  \citenamefont {Yocky},\ and\ \citenamefont {Galayda}}]{EmAk-nphoto-2010}%
  \BibitemOpen
  \bibfield  {author} {\bibinfo {author} {\bibfnamefont {P.}~\bibnamefont
  {Emma}}, \bibinfo {author} {\bibfnamefont {R.}~\bibnamefont {Akre}}, \bibinfo
  {author} {\bibfnamefont {J.}~\bibnamefont {Arthur}},  \emph {et~al.},\
  }\href{http://dx.doi.org/10.1038/nphoton.2010.176} {\bibfield  {journal}
  {\bibinfo  {journal} {Nat Phot.}\ }\textbf {\bibinfo {volume} {4}},\ \bibinfo
  {pages} {641} (\bibinfo {year} {2010})}\BibitemShut {NoStop}%
\bibitem [{\citenamefont {Ishikawa}\ \emph {et~al.}(2012)\citenamefont
  {Ishikawa}, \citenamefont {Aoyagi}, \citenamefont {Asaka}, \citenamefont
  {Asano}, \citenamefont {Azumi}, \citenamefont {Bizen}, \citenamefont {Ego},
  \citenamefont {Fukami}, \citenamefont {Fukui}, \citenamefont {Furukawa},
  \citenamefont {Goto}, \citenamefont {Hanaki}, \citenamefont {Hara},
  \citenamefont {Hasegawa}, \citenamefont {Hatsui}, \citenamefont {Higashiya},
  \citenamefont {Hirono}, \citenamefont {Hosoda}, \citenamefont {Ishii},
  \citenamefont {Inagaki}, \citenamefont {Inubushi}, \citenamefont {Itoga},
  \citenamefont {Joti}, \citenamefont {Kago}, \citenamefont {Kameshima},
  \citenamefont {Kimura}, \citenamefont {Kirihara}, \citenamefont {Kiyomichi},
  \citenamefont {Kobayashi}, \citenamefont {Kondo}, \citenamefont {Kudo},
  \citenamefont {Maesaka}, \citenamefont {Marechal}, \citenamefont {Masuda},
  \citenamefont {Matsubara}, \citenamefont {Matsumoto}, \citenamefont
  {Matsushita}, \citenamefont {Matsui}, \citenamefont {Nagasono}, \citenamefont
  {Nariyama}, \citenamefont {Ohashi}, \citenamefont {Ohata}, \citenamefont
  {Ohshima}, \citenamefont {Ono}, \citenamefont {Otake}, \citenamefont {Saji},
  \citenamefont {Sakurai}, \citenamefont {Sato}, \citenamefont {Sawada},
  \citenamefont {Seike}, \citenamefont {Shirasawa}, \citenamefont {Sugimoto},
  \citenamefont {Suzuki}, \citenamefont {Takahashi}, \citenamefont {Takebe},
  \citenamefont {Takeshita}, \citenamefont {Tamasaku}, \citenamefont {Tanaka},
  \citenamefont {Tanaka}, \citenamefont {Tanaka}, \citenamefont {Togashi},
  \citenamefont {Togawa}, \citenamefont {Tokuhisa}, \citenamefont {Tomizawa},
  \citenamefont {Tono}, \citenamefont {Wu}, \citenamefont {Yabashi},
  \citenamefont {Yamaga}, \citenamefont {Yamashita}, \citenamefont {Yanagida},
  \citenamefont {Zhang}, \citenamefont {Shintake}, \citenamefont {Kitamura},\
  and\ \citenamefont {Kumagai}}]{IsAo-nphoto-2012}%
  \BibitemOpen
  \bibfield  {author} {\bibinfo {author} {\bibfnamefont {T.}~\bibnamefont
  {Ishikawa}}, \bibinfo {author} {\bibfnamefont {H.}~\bibnamefont {Aoyagi}},
  \bibinfo {author} {\bibfnamefont {T.}~\bibnamefont {Asaka}},  \emph
  {et~al.},\ }\href{http://dx.doi.org/10.1038/nphoton.2012.141} {\bibfield
  {journal} {\bibinfo  {journal} {Nat Phot.}\ }\textbf {\bibinfo {volume}
  {6}},\ \bibinfo {pages} {540} (\bibinfo {year} {2012})}\BibitemShut {NoStop}%
\bibitem [{\citenamefont {Allaria}\ \emph {et~al.}(2012)\citenamefont
  {Allaria}, \citenamefont {Appio}, \citenamefont {Badano}, \citenamefont
  {Barletta}, \citenamefont {Bassanese}, \citenamefont {Biedron}, \citenamefont
  {Borga}, \citenamefont {Busetto}, \citenamefont {Castronovo}, \citenamefont
  {Cinquegrana}, \citenamefont {Cleva}, \citenamefont {Cocco}, \citenamefont
  {Cornacchia}, \citenamefont {Craievich}, \citenamefont {Cudin}, \citenamefont
  {D'Auria}, \citenamefont {Dal~Forno}, \citenamefont {Danailov}, \citenamefont
  {De~Monte}, \citenamefont {De~Ninno}, \citenamefont {Delgiusto},
  \citenamefont {Demidovich}, \citenamefont {Di~Mitri}, \citenamefont
  {Diviacco}, \citenamefont {Fabris}, \citenamefont {Fabris}, \citenamefont
  {Fawley}, \citenamefont {Ferianis}, \citenamefont {Ferrari}, \citenamefont
  {Ferry}, \citenamefont {Froehlich}, \citenamefont {Furlan}, \citenamefont
  {Gaio}, \citenamefont {Gelmetti}, \citenamefont {Giannessi}, \citenamefont
  {Giannini}, \citenamefont {Gobessi}, \citenamefont {Ivanov}, \citenamefont
  {Karantzoulis}, \citenamefont {Lonza}, \citenamefont {Lutman}, \citenamefont
  {Mahieu}, \citenamefont {Milloch}, \citenamefont {Milton}, \citenamefont
  {Musardo}, \citenamefont {Nikolov}, \citenamefont {Noe}, \citenamefont
  {Parmigiani}, \citenamefont {Penco}, \citenamefont {Petronio}, \citenamefont
  {Pivetta}, \citenamefont {Predonzani}, \citenamefont {Rossi}, \citenamefont
  {Rumiz}, \citenamefont {Salom}, \citenamefont {Scafuri}, \citenamefont
  {Serpico}, \citenamefont {Sigalotti}, \citenamefont {Spampinati},
  \citenamefont {Spezzani}, \citenamefont {Svandrlik}, \citenamefont {Svetina},
  \citenamefont {Tazzari}, \citenamefont {Trovo}, \citenamefont {Umer},
  \citenamefont {Vascotto}, \citenamefont {Veronese}, \citenamefont
  {Visintini}, \citenamefont {Zaccaria}, \citenamefont {Zangrando},\ and\
  \citenamefont {Zangrando}}]{MuNi-NatPhoton-2012}%
  \BibitemOpen
  \bibfield  {author} {\bibinfo {author} {\bibfnamefont {E.}~\bibnamefont
  {Allaria}}, \bibinfo {author} {\bibfnamefont {R.}~\bibnamefont {Appio}},
  \bibinfo {author} {\bibfnamefont {L.}~\bibnamefont {Badano}},  \emph
  {et~al.},\ }\href{\doibase 10.1038/NPHOTON.2012.233} {\bibfield  {journal}
  {\bibinfo  {journal} {Nat. Phot.}\ }\textbf {\bibinfo {volume} {6}},\
  \bibinfo {pages} {699} (\bibinfo {year} {2012})}\BibitemShut {NoStop}%
\bibitem [{\citenamefont {Young}\ \emph {et~al.}(2010)\citenamefont {Young},
  \citenamefont {Kanter}, \citenamefont {Kr\"assig}, \citenamefont {Li},
  \citenamefont {March}, \citenamefont {Pratt}, \citenamefont {Santra},
  \citenamefont {Southworth}, \citenamefont {Rohringer}, \citenamefont
  {{DiMauro}}, \citenamefont {Doumy}, \citenamefont {Roedig}, \citenamefont
  {Berrah}, \citenamefont {Fang}, \citenamefont {Hoener}, \citenamefont
  {Bucksbaum}, \citenamefont {Cryan}, \citenamefont {Ghimire}, \citenamefont
  {Glownia}, \citenamefont {Reis}, \citenamefont {Bozek}, \citenamefont
  {Bostedt},\ and\ \citenamefont {Messerschmidt}}]{YoKa-Nature-2010}%
  \BibitemOpen
  \bibfield  {author} {\bibinfo {author} {\bibfnamefont {L.}~\bibnamefont
  {Young}}, \bibinfo {author} {\bibfnamefont {E.~P.}\ \bibnamefont {Kanter}},
  \bibinfo {author} {\bibfnamefont {B.}~\bibnamefont {Kr\"assig}},  \emph
  {et~al.},\ }\href{\doibase 10.1038/nature09177} {\bibfield  {journal}
  {\bibinfo  {journal} {Nature}\ }\textbf {\bibinfo {volume} {466}},\ \bibinfo
  {pages} {56} (\bibinfo {year} {2010})}\BibitemShut {NoStop}%
\bibitem [{\citenamefont {Doumy}\ \emph {et~al.}(2011)\citenamefont {Doumy},
  \citenamefont {Roedig}, \citenamefont {Son}, \citenamefont {Blaga},
  \citenamefont {DiChiara}, \citenamefont {Santra}, \citenamefont {Berrah},
  \citenamefont {Bostedt}, \citenamefont {Bozek}, \citenamefont {Bucksbaum},
  \citenamefont {Cryan}, \citenamefont {Fang}, \citenamefont {Ghimire},
  \citenamefont {Glownia}, \citenamefont {Hoener}, \citenamefont {Kanter},
  \citenamefont {Kr\"assig}, \citenamefont {Kuebel}, \citenamefont
  {Messerschmidt}, \citenamefont {Paulus}, \citenamefont {Reis}, \citenamefont
  {Rohringer}, \citenamefont {Young}, \citenamefont {Agostini},\ and\
  \citenamefont {DiMauro}}]{DoRo-PRL-2011}%
  \BibitemOpen
  \bibfield  {author} {\bibinfo {author} {\bibfnamefont {G.}~\bibnamefont
  {Doumy}}, \bibinfo {author} {\bibfnamefont {C.}~\bibnamefont {Roedig}},
  \bibinfo {author} {\bibfnamefont {S.-K.}\ \bibnamefont {Son}},  \emph
  {et~al.},\ }\href{\doibase 10.1103/PhysRevLett.106.083002} {\bibfield
  {journal} {\bibinfo  {journal} {Phys. Rev. Lett.}\ }\textbf {\bibinfo
  {volume} {106}},\ \bibinfo {pages} {083002} (\bibinfo {year}
  {2011})}\BibitemShut {NoStop}%
\bibitem [{\citenamefont {Rohringer}\ \emph {et~al.}(2012)\citenamefont
  {Rohringer}, \citenamefont {Ryan}, \citenamefont {London}, \citenamefont
  {Purvis}, \citenamefont {Albert}, \citenamefont {Dunn}, \citenamefont
  {Bozek}, \citenamefont {Bostedt}, \citenamefont {Graf}, \citenamefont {Hill},
  \citenamefont {Hau-Riege},\ and\ \citenamefont {Rocca}}]{RoRy-Nature-2012}%
  \BibitemOpen
  \bibfield  {author} {\bibinfo {author} {\bibfnamefont {N.}~\bibnamefont
  {Rohringer}}, \bibinfo {author} {\bibfnamefont {D.}~\bibnamefont {Ryan}},
  \bibinfo {author} {\bibfnamefont {R.~A.}\ \bibnamefont {London}},  \emph
  {et~al.},\ }\href{\doibase 10.1038/nature10721} {\bibfield  {journal}
  {\bibinfo  {journal} {Nature}\ }\textbf {\bibinfo {volume} {481}},\ \bibinfo
  {pages} {488} (\bibinfo {year} {2012})}\BibitemShut {NoStop}%
\bibitem [{\citenamefont {Hoener}\ \emph {et~al.}(2010)\citenamefont {Hoener},
  \citenamefont {Fang}, \citenamefont {Kornilov}, \citenamefont {Gessner},
  \citenamefont {Pratt}, \citenamefont {G\"uhr}, \citenamefont {Kanter},
  \citenamefont {Blaga}, \citenamefont {Bostedt}, \citenamefont {Bozek},
  \citenamefont {Bucksbaum}, \citenamefont {Buth}, \citenamefont {Chen},
  \citenamefont {Coffee}, \citenamefont {Cryan}, \citenamefont {DiMauro},
  \citenamefont {Glownia}, \citenamefont {Hosler}, \citenamefont {Kukk},
  \citenamefont {Leone}, \citenamefont {McFarland}, \citenamefont
  {Messerschmidt}, \citenamefont {Murphy}, \citenamefont {Petrovic},
  \citenamefont {Rolles},\ and\ \citenamefont {Berrah}}]{HoFa-PRL-2010}%
  \BibitemOpen
  \bibfield  {author} {\bibinfo {author} {\bibfnamefont {M.}~\bibnamefont
  {Hoener}}, \bibinfo {author} {\bibfnamefont {L.}~\bibnamefont {Fang}},
  \bibinfo {author} {\bibfnamefont {O.}~\bibnamefont {Kornilov}},  \emph
  {et~al.},\ }\href{\doibase 10.1103/PhysRevLett.104.253002} {\bibfield
  {journal} {\bibinfo  {journal} {Phys. Rev. Lett.}\ }\textbf {\bibinfo
  {volume} {104}},\ \bibinfo {pages} {253002} (\bibinfo {year}
  {2010})}\BibitemShut {NoStop}%
\bibitem [{\citenamefont {Cryan}\ \emph {et~al.}(2010)\citenamefont {Cryan},
  \citenamefont {Glownia}, \citenamefont {Andreasson}, \citenamefont
  {Belkacem}, \citenamefont {Berrah}, \citenamefont {Blaga}, \citenamefont
  {Bostedt}, \citenamefont {Bozek}, \citenamefont {Buth}, \citenamefont
  {DiMauro}, \citenamefont {Fang}, \citenamefont {Gessner}, \citenamefont
  {Guehr}, \citenamefont {Hajdu}, \citenamefont {Hertlein}, \citenamefont
  {Hoener}, \citenamefont {Kornilov}, \citenamefont {Marangos}, \citenamefont
  {March}, \citenamefont {McFarland}, \citenamefont {Merdji}, \citenamefont
  {Petrovi\ifmmode~\acute{c}\else \'{c}\fi{}}, \citenamefont {Raman},
  \citenamefont {Ray}, \citenamefont {Reis}, \citenamefont {Tarantelli},
  \citenamefont {Trigo}, \citenamefont {White}, \citenamefont {White},
  \citenamefont {Young}, \citenamefont {Bucksbaum},\ and\ \citenamefont
  {Coffee}}]{CrGl-PRL-2010}%
  \BibitemOpen
  \bibfield  {author} {\bibinfo {author} {\bibfnamefont {J.~P.}\ \bibnamefont
  {Cryan}}, \bibinfo {author} {\bibfnamefont {J.~M.}\ \bibnamefont {Glownia}},
  \bibinfo {author} {\bibfnamefont {J.}~\bibnamefont {Andreasson}},  \emph
  {et~al.},\ }\href{\doibase 10.1103/PhysRevLett.105.083004} {\bibfield
  {journal} {\bibinfo  {journal} {Phys. Rev. Lett.}\ }\textbf {\bibinfo
  {volume} {105}},\ \bibinfo {pages} {083004} (\bibinfo {year}
  {2010})}\BibitemShut {NoStop}%
\bibitem [{\citenamefont {Bostedt}\ \emph {et~al.}(2010)\citenamefont
  {Bostedt}, \citenamefont {Thomas}, \citenamefont {Hoener}, \citenamefont
  {M\"oller}, \citenamefont {Saalmann}, \citenamefont {Georgescu},
  \citenamefont {Gnodtke},\ and\ \citenamefont {Rost}}]{BoTh-NJP-2010}%
  \BibitemOpen
  \bibfield  {author} {\bibinfo {author} {\bibfnamefont {C.}~\bibnamefont
  {Bostedt}}, \bibinfo {author} {\bibfnamefont {H.}~\bibnamefont {Thomas}},
  \bibinfo {author} {\bibfnamefont {M.}~\bibnamefont {Hoener}},  \emph
  {et~al.},\ }\href{http://stacks.iop.org/1367-2630/12/i=8/a=083004} {\bibfield
   {journal} {\bibinfo  {journal} {New J. Phys.}\ }\textbf {\bibinfo {volume}
  {12}},\ \bibinfo {pages} {083004} (\bibinfo {year} {2010})}\BibitemShut
  {NoStop}%
\bibitem [{\citenamefont {Thomas}\ \emph {et~al.}(2012)\citenamefont {Thomas},
  \citenamefont {Helal}, \citenamefont {Hoffmann}, \citenamefont {Kandadai},
  \citenamefont {Keto}, \citenamefont {Andreasson}, \citenamefont {Iwan},
  \citenamefont {Seibert}, \citenamefont {Timneanu}, \citenamefont {Hajdu},
  \citenamefont {Adolph}, \citenamefont {Gorkhover}, \citenamefont {Rupp},
  \citenamefont {Schorb}, \citenamefont {M\"oller}, \citenamefont {Doumy},
  \citenamefont {DiMauro}, \citenamefont {Hoener}, \citenamefont {Murphy},
  \citenamefont {Berrah}, \citenamefont {Messerschmidt}, \citenamefont {Bozek},
  \citenamefont {Bostedt},\ and\ \citenamefont {Ditmire}}]{ThHe-PRL-2012}%
  \BibitemOpen
  \bibfield  {author} {\bibinfo {author} {\bibfnamefont {H.}~\bibnamefont
  {Thomas}}, \bibinfo {author} {\bibfnamefont {A.}~\bibnamefont {Helal}},
  \bibinfo {author} {\bibfnamefont {K.}~\bibnamefont {Hoffmann}},  \emph
  {et~al.},\ }\href{\doibase 10.1103/PhysRevLett.108.133401} {\bibfield
  {journal} {\bibinfo  {journal} {Phys. Rev. Lett.}\ }\textbf {\bibinfo
  {volume} {108}},\ \bibinfo {pages} {133401} (\bibinfo {year}
  {2012})}\BibitemShut {NoStop}%
\bibitem [{\citenamefont {Scully}\ and\ \citenamefont
  {Zubairy}(1997)}]{Scully-book}%
  \BibitemOpen
  \bibfield  {author} {\bibinfo {author} {\bibfnamefont {M.~O.}\ \bibnamefont
  {Scully}}\ and\ \bibinfo {author} {\bibfnamefont {M.~S.}\ \bibnamefont
  {Zubairy}},\ }\href@noop {} {\emph {\bibinfo {title} {Quantum Optics}}}\
  (\bibinfo  {publisher} {Cambridge University Press, Cambridge, UK},\ \bibinfo
  {year} {1997})\BibitemShut {NoStop}%
\bibitem [{\citenamefont {Wu}\ \emph {et~al.}(1977)\citenamefont {Wu},
  \citenamefont {Ezekiel}, \citenamefont {Ducloy},\ and\ \citenamefont
  {Mollow}}]{WuEz-PRL-1977}%
  \BibitemOpen
  \bibfield  {author} {\bibinfo {author} {\bibfnamefont {F.~Y.}\ \bibnamefont
  {Wu}}, \bibinfo {author} {\bibfnamefont {S.}~\bibnamefont {Ezekiel}},
  \bibinfo {author} {\bibfnamefont {M.}~\bibnamefont {Ducloy}}, \ and\ \bibinfo
  {author} {\bibfnamefont {B.~R.}\ \bibnamefont {Mollow}},\ }\href{\doibase
  10.1103/PhysRevLett.38.1077} {\bibfield  {journal} {\bibinfo  {journal}
  {Phys. Rev. Lett.}\ }\textbf {\bibinfo {volume} {38}},\ \bibinfo {pages}
  {1077} (\bibinfo {year} {1977})}\BibitemShut {NoStop}%
\bibitem [{\citenamefont {Sellin}\ \emph {et~al.}(1996)\citenamefont {Sellin},
  \citenamefont {Wilson}, \citenamefont {Meduri},\ and\ \citenamefont
  {Mossberg}}]{SeWi-PRA-1996}%
  \BibitemOpen
  \bibfield  {author} {\bibinfo {author} {\bibfnamefont {P.~B.}\ \bibnamefont
  {Sellin}}, \bibinfo {author} {\bibfnamefont {G.~A.}\ \bibnamefont {Wilson}},
  \bibinfo {author} {\bibfnamefont {K.~K.}\ \bibnamefont {Meduri}}, \ and\
  \bibinfo {author} {\bibfnamefont {T.~W.}\ \bibnamefont {Mossberg}},\
  }\href{\doibase 10.1103/PhysRevA.54.2402} {\bibfield  {journal} {\bibinfo
  {journal} {Phys. Rev. A}\ }\textbf {\bibinfo {volume} {54}},\ \bibinfo
  {pages} {2402} (\bibinfo {year} {1996})}\BibitemShut {NoStop}%
\bibitem [{\citenamefont {Kanter}\ \emph {et~al.}(2011)\citenamefont {Kanter},
  \citenamefont {Kr\"assig}, \citenamefont {Li}, \citenamefont {March},
  \citenamefont {Ho}, \citenamefont {Rohringer}, \citenamefont {Santra},
  \citenamefont {Southworth}, \citenamefont {DiMauro}, \citenamefont {Doumy},
  \citenamefont {Roedig}, \citenamefont {Berrah}, \citenamefont {Fang},
  \citenamefont {Hoener}, \citenamefont {Bucksbaum}, \citenamefont {Ghimire},
  \citenamefont {Reis}, \citenamefont {Bozek}, \citenamefont {Bostedt},
  \citenamefont {Messerschmidt},\ and\ \citenamefont {Young}}]{KaKr-PRL-2011}%
  \BibitemOpen
  \bibfield  {author} {\bibinfo {author} {\bibfnamefont {E.~P.}\ \bibnamefont
  {Kanter}}, \bibinfo {author} {\bibfnamefont {B.}~\bibnamefont {Kr\"assig}},
  \bibinfo {author} {\bibfnamefont {Y.}~\bibnamefont {Li}},  \emph {et~al.},\
  }\href{\doibase 10.1103/PhysRevLett.107.233001} {\bibfield  {journal}
  {\bibinfo  {journal} {Phys. Rev. Lett.}\ }\textbf {\bibinfo {volume} {107}},\
  \bibinfo {pages} {233001} (\bibinfo {year} {2011})}\BibitemShut {NoStop}%
\bibitem [{\citenamefont {Rohringer}\ and\ \citenamefont
  {Santra}(2012)}]{RoSa-PRA-2012}%
  \BibitemOpen
  \bibfield  {author} {\bibinfo {author} {\bibfnamefont {N.}~\bibnamefont
  {Rohringer}}\ and\ \bibinfo {author} {\bibfnamefont {R.}~\bibnamefont
  {Santra}},\ }\href{\doibase 10.1103/PhysRevA.86.043434} {\bibfield  {journal}
  {\bibinfo  {journal} {Phys. Rev. A}\ }\textbf {\bibinfo {volume} {86}},\
  \bibinfo {pages} {043434} (\bibinfo {year} {2012})}\BibitemShut {NoStop}%
\bibitem [{\citenamefont {Buth}\ \emph {et~al.}(2007)\citenamefont {Buth},
  \citenamefont {Santra},\ and\ \citenamefont {Young}}]{BuSa-PRL98-2007}%
  \BibitemOpen
  \bibfield  {author} {\bibinfo {author} {\bibfnamefont {C.}~\bibnamefont
  {Buth}}, \bibinfo {author} {\bibfnamefont {R.}~\bibnamefont {Santra}}, \ and\
  \bibinfo {author} {\bibfnamefont {L.}~\bibnamefont {Young}},\ }\href{\doibase
  10.1103/PhysRevLett.98.253001} {\bibfield  {journal} {\bibinfo  {journal}
  {Phys. Rev. Lett.}\ }\textbf {\bibinfo {volume} {98}},\ \bibinfo {pages}
  {253001} (\bibinfo {year} {2007})}\BibitemShut {NoStop}%
\bibitem [{\citenamefont {Glover}\ \emph {et~al.}(2010)\citenamefont {Glover},
  \citenamefont {Hertlein}, \citenamefont {Southworth}, \citenamefont
  {Allison}, \citenamefont {van Tilborg}, \citenamefont {Kanter}, \citenamefont
  {Krassig}, \citenamefont {Varma}, \citenamefont {Rude}, \citenamefont
  {Santra}, \citenamefont {Belkacem},\ and\ \citenamefont
  {Young}}]{GlHe-NatPhys-2010}%
  \BibitemOpen
  \bibfield  {author} {\bibinfo {author} {\bibfnamefont {T.~E.}\ \bibnamefont
  {Glover}}, \bibinfo {author} {\bibfnamefont {M.~P.}\ \bibnamefont
  {Hertlein}}, \bibinfo {author} {\bibfnamefont {S.~H.}\ \bibnamefont
  {Southworth}},  \emph {et~al.},\ }\href{\doibase 10.1038/nphys1430}
  {\bibfield  {journal} {\bibinfo  {journal} {Nat. Phys.}\ }\textbf {\bibinfo
  {volume} {6}},\ \bibinfo {pages} {69} (\bibinfo {year} {2010})}\BibitemShut
  {NoStop}%
\bibitem [{\citenamefont {Thompson}\ \emph {et~al.}()\citenamefont {Thompson},
  \citenamefont {Attwood}, \citenamefont {Gullikson}, \citenamefont {Howells},
  \citenamefont {Kortright}, \citenamefont {Robinson}, \citenamefont
  {Underwood}, \citenamefont {Kim}, \citenamefont {Kirz}, \citenamefont
  {Lindau}, \citenamefont {Pianetta}, \citenamefont {Winick}, \citenamefont
  {Williams},\ and\ \citenamefont {Scofield}}]{XDB}%
  \BibitemOpen
  \bibfield  {author} {\bibinfo {author} {\bibfnamefont {A.~C.}\ \bibnamefont
  {Thompson}}, \bibinfo {author} {\bibfnamefont {D.~T.}\ \bibnamefont
  {Attwood}}, \bibinfo {author} {\bibfnamefont {E.~M.}\ \bibnamefont
  {Gullikson}},  \emph {et~al.},\ }\href@noop {} {}\bibinfo {note} {X-Ray Data
  Booklet, \url{http://xdb.lbl.gov}}\BibitemShut {NoStop}%
\bibitem [{\citenamefont {Cavaletto}\ \emph {et~al.}(2012)\citenamefont
  {Cavaletto}, \citenamefont {Buth}, \citenamefont {Harman}, \citenamefont
  {Kanter}, \citenamefont {Southworth}, \citenamefont {Young},\ and\
  \citenamefont {Keitel}}]{CaKe-PRA-2012}%
  \BibitemOpen
  \bibfield  {author} {\bibinfo {author} {\bibfnamefont {S.~M.}\ \bibnamefont
  {Cavaletto}}, \bibinfo {author} {\bibfnamefont {C.}~\bibnamefont {Buth}},
  \bibinfo {author} {\bibfnamefont {Z.}~\bibnamefont {Harman}}, \bibinfo
  {author} {\bibfnamefont {E.~P.}\ \bibnamefont {Kanter}}, \bibinfo {author}
  {\bibfnamefont {S.~H.}\ \bibnamefont {Southworth}}, \bibinfo {author}
  {\bibfnamefont {L.}~\bibnamefont {Young}}, \ and\ \bibinfo {author}
  {\bibfnamefont {C.~H.}\ \bibnamefont {Keitel}},\ }\href{\doibase
  10.1103/PhysRevA.86.033402} {\bibfield  {journal} {\bibinfo  {journal} {Phys.
  Rev. A}\ }\textbf {\bibinfo {volume} {86}},\ \bibinfo {pages} {033402}
  (\bibinfo {year} {2012})}\BibitemShut {NoStop}%
\bibitem [{\citenamefont {Rohringer}\ and\ \citenamefont
  {Santra}(2008)}]{RoSa-PRA-2008}%
  \BibitemOpen
  \bibfield  {author} {\bibinfo {author} {\bibfnamefont {N.}~\bibnamefont
  {Rohringer}}\ and\ \bibinfo {author} {\bibfnamefont {R.}~\bibnamefont
  {Santra}},\ }\href{\doibase 10.1103/PhysRevA.77.053404} {\bibfield  {journal}
  {\bibinfo  {journal} {Phys. Rev. A}\ }\textbf {\bibinfo {volume} {77}},\
  \bibinfo {pages} {053404} (\bibinfo {year} {2008})}\BibitemShut {NoStop}%
\bibitem [{\citenamefont {Liu}\ \emph {et~al.}(2010)\citenamefont {Liu},
  \citenamefont {Sun}, \citenamefont {Wang}, \citenamefont {\AA{}gren},\ and\
  \citenamefont {Gel'mukhanov}}]{CaGe-PRA-2010}%
  \BibitemOpen
  \bibfield  {author} {\bibinfo {author} {\bibfnamefont {J.-C.}\ \bibnamefont
  {Liu}}, \bibinfo {author} {\bibfnamefont {Y.-P.}\ \bibnamefont {Sun}},
  \bibinfo {author} {\bibfnamefont {C.-K.}\ \bibnamefont {Wang}}, \bibinfo
  {author} {\bibfnamefont {H.}~\bibnamefont {\AA{}gren}}, \ and\ \bibinfo
  {author} {\bibfnamefont {F.}~\bibnamefont {Gel'mukhanov}},\ }\href{\doibase
  10.1103/PhysRevA.81.043412} {\bibfield  {journal} {\bibinfo  {journal} {Phys.
  Rev. A}\ }\textbf {\bibinfo {volume} {81}},\ \bibinfo {pages} {043412}
  (\bibinfo {year} {2010})}\BibitemShut {NoStop}%
\bibitem [{\citenamefont {Nikolopoulos}\ \emph {et~al.}(2011)\citenamefont
  {Nikolopoulos}, \citenamefont {Kelly},\ and\ \citenamefont
  {Costello}}]{NiCo-PRA-2011}%
  \BibitemOpen
  \bibfield  {author} {\bibinfo {author} {\bibfnamefont {L.~A.~A.}\
  \bibnamefont {Nikolopoulos}}, \bibinfo {author} {\bibfnamefont {T.~J.}\
  \bibnamefont {Kelly}}, \ and\ \bibinfo {author} {\bibfnamefont {J.~T.}\
  \bibnamefont {Costello}},\ }\href{\doibase 10.1103/PhysRevA.84.063419}
  {\bibfield  {journal} {\bibinfo  {journal} {Phys. Rev. A}\ }\textbf {\bibinfo
  {volume} {84}},\ \bibinfo {pages} {063419} (\bibinfo {year}
  {2011})}\BibitemShut {NoStop}%
\bibitem [{\citenamefont {Sako}\ \emph {et~al.}(2011)\citenamefont {Sako},
  \citenamefont {Adachi}, \citenamefont {Yagishita}, \citenamefont {Yabashi},
  \citenamefont {Tanaka}, \citenamefont {Nagasono},\ and\ \citenamefont
  {Ishikawa}}]{SaAd-PRA-2011}%
  \BibitemOpen
  \bibfield  {author} {\bibinfo {author} {\bibfnamefont {T.}~\bibnamefont
  {Sako}}, \bibinfo {author} {\bibfnamefont {J.}~\bibnamefont {Adachi}},
  \bibinfo {author} {\bibfnamefont {A.}~\bibnamefont {Yagishita}}, \bibinfo
  {author} {\bibfnamefont {M.}~\bibnamefont {Yabashi}}, \bibinfo {author}
  {\bibfnamefont {T.}~\bibnamefont {Tanaka}}, \bibinfo {author} {\bibfnamefont
  {M.}~\bibnamefont {Nagasono}}, \ and\ \bibinfo {author} {\bibfnamefont
  {T.}~\bibnamefont {Ishikawa}},\ }\href{\doibase 10.1103/PhysRevA.84.053419}
  {\bibfield  {journal} {\bibinfo  {journal} {Phys. Rev. A}\ }\textbf {\bibinfo
  {volume} {84}},\ \bibinfo {pages} {053419} (\bibinfo {year}
  {2011})}\BibitemShut {NoStop}%
\bibitem [{\citenamefont {Haxton}\ and\ \citenamefont
  {McCurdy}(2014)}]{HaMc-PRA-2014}%
  \BibitemOpen
  \bibfield  {author} {\bibinfo {author} {\bibfnamefont {D.~J.}\ \bibnamefont
  {Haxton}}\ and\ \bibinfo {author} {\bibfnamefont {C.~W.}\ \bibnamefont
  {McCurdy}},\ }\href{\doibase 10.1103/PhysRevA.90.053426} {\bibfield
  {journal} {\bibinfo  {journal} {Phys. Rev. A}\ }\textbf {\bibinfo {volume}
  {90}},\ \bibinfo {pages} {053426} (\bibinfo {year} {2014})}\BibitemShut
  {NoStop}%
\bibitem [{\citenamefont {Demekhin}\ \emph {et~al.}(2011)\citenamefont
  {Demekhin}, \citenamefont {Chiang},\ and\ \citenamefont
  {Cederbaum}}]{DeCh-PRA-2011}%
  \BibitemOpen
  \bibfield  {author} {\bibinfo {author} {\bibfnamefont {P.~V.}\ \bibnamefont
  {Demekhin}}, \bibinfo {author} {\bibfnamefont {Y.-C.}\ \bibnamefont
  {Chiang}}, \ and\ \bibinfo {author} {\bibfnamefont {L.~S.}\ \bibnamefont
  {Cederbaum}},\ }\href{\doibase 10.1103/PhysRevA.84.033417} {\bibfield
  {journal} {\bibinfo  {journal} {Phys. Rev. A}\ }\textbf {\bibinfo {volume}
  {84}},\ \bibinfo {pages} {033417} (\bibinfo {year} {2011})}\BibitemShut
  {NoStop}%
\bibitem [{\citenamefont {M\"uller}\ and\ \citenamefont
  {Demekhin}(2015)}]{MuDe-JPB-2015}%
  \BibitemOpen
  \bibfield  {author} {\bibinfo {author} {\bibfnamefont {A.~D.}\ \bibnamefont
  {M\"uller}}\ and\ \bibinfo {author} {\bibfnamefont {P.~V.}\ \bibnamefont
  {Demekhin}},\ }\href{http://stacks.iop.org/0953-4075/48/i=7/a=075602}
  {\bibfield  {journal} {\bibinfo  {journal} {J. Phys. B: At., Mol. Opt.
  Phys.}\ }\textbf {\bibinfo {volume} {48}},\ \bibinfo {pages} {075602}
  (\bibinfo {year} {2015})}\BibitemShut {NoStop}%
\bibitem [{\citenamefont {Chatterjee}\ and\ \citenamefont
  {Nakajima}(2015)}]{ChNa-PRA-2015}%
  \BibitemOpen
  \bibfield  {author} {\bibinfo {author} {\bibfnamefont {S.}~\bibnamefont
  {Chatterjee}}\ and\ \bibinfo {author} {\bibfnamefont {T.}~\bibnamefont
  {Nakajima}},\ }\href{\doibase 10.1103/PhysRevA.91.043413} {\bibfield
  {journal} {\bibinfo  {journal} {Phys. Rev. A}\ }\textbf {\bibinfo {volume}
  {91}},\ \bibinfo {pages} {043413} (\bibinfo {year} {2015})}\BibitemShut
  {NoStop}%
\bibitem [{\citenamefont {Fano}\ and\ \citenamefont
  {Cooper}(1968)}]{FaCo-RMP-1968}%
  \BibitemOpen
  \bibfield  {author} {\bibinfo {author} {\bibfnamefont {U.}~\bibnamefont
  {Fano}}\ and\ \bibinfo {author} {\bibfnamefont {J.~W.}\ \bibnamefont
  {Cooper}},\ }\href{\doibase 10.1103/RevModPhys.40.441} {\bibfield  {journal}
  {\bibinfo  {journal} {Rev. Mod. Phys.}\ }\textbf {\bibinfo {volume} {40}},\
  \bibinfo {pages} {441} (\bibinfo {year} {1968})}\BibitemShut {NoStop}%
\bibitem [{\citenamefont {Starace}(1982)}]{St-Springer-1980}%
  \BibitemOpen
  \bibfield  {author} {\bibinfo {author} {\bibfnamefont {A.~F.}\ \bibnamefont
  {Starace}},\ }in\
  \href{http://www.springer.com/materials/book/978-3-642-46455-3} {\emph
  {\bibinfo {booktitle} {Encyclopedia of Physics}}},\ Vol.\ \bibinfo {volume}
  {31: Corpuscles and Radiation in Matter I},\ \bibinfo {editor} {edited by\
  \bibinfo {editor} {\bibfnamefont {W.}~\bibnamefont {Mehlhorn}}}\ (\bibinfo
  {publisher} {Springer, Berlin},\ \bibinfo {year} {1982})\ Chap.\ \bibinfo
  {chapter} {Theory of Atomic Photoionization}, pp.\ \bibinfo {pages}
  {1--121}\BibitemShut {NoStop}%
\bibitem [{\citenamefont {Krebs}\ \emph {et~al.}(2014)\citenamefont {Krebs},
  \citenamefont {Pabst},\ and\ \citenamefont {Santra}}]{KrPa-AJP-2014}%
  \BibitemOpen
  \bibfield  {author} {\bibinfo {author} {\bibfnamefont {D.}~\bibnamefont
  {Krebs}}, \bibinfo {author} {\bibfnamefont {S.}~\bibnamefont {Pabst}}, \ and\
  \bibinfo {author} {\bibfnamefont {R.}~\bibnamefont {Santra}},\
  }\href{\doibase http://dx.doi.org/10.1119/1.4827015} {\bibfield  {journal}
  {\bibinfo  {journal} {Am. J. Phys.}\ }\textbf {\bibinfo {volume} {82}},\
  \bibinfo {pages} {113} (\bibinfo {year} {2014})}\BibitemShut {NoStop}%
\bibitem [{\citenamefont {Frolov}\ \emph {et~al.}(2009)\citenamefont {Frolov},
  \citenamefont {Manakov}, \citenamefont {Sarantseva}, \citenamefont {Emelin},
  \citenamefont {Ryabikin},\ and\ \citenamefont {Starace}}]{FrSt-PRL-2009}%
  \BibitemOpen
  \bibfield  {author} {\bibinfo {author} {\bibfnamefont {M.~V.}\ \bibnamefont
  {Frolov}}, \bibinfo {author} {\bibfnamefont {N.~L.}\ \bibnamefont {Manakov}},
  \bibinfo {author} {\bibfnamefont {T.~S.}\ \bibnamefont {Sarantseva}},
  \bibinfo {author} {\bibfnamefont {M.~Y.}\ \bibnamefont {Emelin}}, \bibinfo
  {author} {\bibfnamefont {M.~Y.}\ \bibnamefont {Ryabikin}}, \ and\ \bibinfo
  {author} {\bibfnamefont {A.~F.}\ \bibnamefont {Starace}},\ }\href{\doibase
  10.1103/PhysRevLett.102.243901} {\bibfield  {journal} {\bibinfo  {journal}
  {Phys. Rev. Lett.}\ }\textbf {\bibinfo {volume} {102}},\ \bibinfo {pages}
  {243901} (\bibinfo {year} {2009})}\BibitemShut {NoStop}%
\bibitem [{\citenamefont {Shiner}\ \emph {et~al.}(2011)\citenamefont {Shiner},
  \citenamefont {Schmidt}, \citenamefont {{Trallero-Herrero}}, \citenamefont
  {W\"orner}, \citenamefont {Patchkovskii}, \citenamefont {Corkum},
  \citenamefont {Kieffer}, \citenamefont {Legare},\ and\ \citenamefont
  {Villeneuve}}]{ShVi-NatPhys-2011}%
  \BibitemOpen
  \bibfield  {author} {\bibinfo {author} {\bibfnamefont {A.~D.}\ \bibnamefont
  {Shiner}}, \bibinfo {author} {\bibfnamefont {B.~E.}\ \bibnamefont {Schmidt}},
  \bibinfo {author} {\bibfnamefont {C.}~\bibnamefont {{Trallero-Herrero}}},
  \emph {et~al.},\ }\href{\doibase 10.1038/nphys1940} {\bibfield  {journal}
  {\bibinfo  {journal} {Nat. Phys.}\ }\textbf {\bibinfo {volume} {7}},\
  \bibinfo {pages} {464} (\bibinfo {year} {2011})}\BibitemShut {NoStop}%
\bibitem [{\citenamefont {Pabst}\ and\ \citenamefont
  {Santra}(2013)}]{PaSa-PRL-2013}%
  \BibitemOpen
  \bibfield  {author} {\bibinfo {author} {\bibfnamefont {S.}~\bibnamefont
  {Pabst}}\ and\ \bibinfo {author} {\bibfnamefont {R.}~\bibnamefont {Santra}},\
  }\href{\doibase 10.1103/PhysRevLett.111.233005} {\bibfield  {journal}
  {\bibinfo  {journal} {Phys. Rev. Lett.}\ }\textbf {\bibinfo {volume} {111}},\
  \bibinfo {pages} {233005} (\bibinfo {year} {2013})}\BibitemShut {NoStop}%
\bibitem [{\citenamefont {Wendin}(1973)}]{We-JPB-1973}%
  \BibitemOpen
  \bibfield  {author} {\bibinfo {author} {\bibfnamefont {G.}~\bibnamefont
  {Wendin}},\ }\href{http://stacks.iop.org/0022-3700/6/i=1/a=007} {\bibfield
  {journal} {\bibinfo  {journal} {J. Phys. B}\ }\textbf {\bibinfo {volume}
  {6}},\ \bibinfo {pages} {42} (\bibinfo {year} {1973})}\BibitemShut {NoStop}%
\bibitem [{\citenamefont {Chen}\ \emph {et~al.}(2015)\citenamefont {Chen},
  \citenamefont {Pabst}, \citenamefont {Karamatskou},\ and\ \citenamefont
  {Santra}}]{ChPa-PRA-2015}%
  \BibitemOpen
  \bibfield  {author} {\bibinfo {author} {\bibfnamefont {Y.-J.}\ \bibnamefont
  {Chen}}, \bibinfo {author} {\bibfnamefont {S.}~\bibnamefont {Pabst}},
  \bibinfo {author} {\bibfnamefont {A.}~\bibnamefont {Karamatskou}}, \ and\
  \bibinfo {author} {\bibfnamefont {R.}~\bibnamefont {Santra}},\
  }\href{\doibase 10.1103/PhysRevA.91.032503} {\bibfield  {journal} {\bibinfo
  {journal} {Phys. Rev. A}\ }\textbf {\bibinfo {volume} {91}},\ \bibinfo
  {pages} {032503} (\bibinfo {year} {2015})}\BibitemShut {NoStop}%
\bibitem [{\citenamefont {Mazza}\ \emph {et~al.}(2015)\citenamefont {Mazza},
  \citenamefont {Karamatskou}, \citenamefont {Ilchen}, \citenamefont
  {Bakhtiarzadeh}, \citenamefont {Rafipoor}, \citenamefont {O/'Keeffe},
  \citenamefont {Kelly}, \citenamefont {Walsh}, \citenamefont {Costello},
  \citenamefont {Meyer},\ and\ \citenamefont {Santra}}]{MaKa-NatComm-2015}%
  \BibitemOpen
  \bibfield  {author} {\bibinfo {author} {\bibfnamefont {T.}~\bibnamefont
  {Mazza}}, \bibinfo {author} {\bibfnamefont {A.}~\bibnamefont {Karamatskou}},
  \bibinfo {author} {\bibfnamefont {M.}~\bibnamefont {Ilchen}},  \emph
  {et~al.},\ }\href{http://dx.doi.org/10.1038/ncomms7799} {\bibfield  {journal}
  {\bibinfo  {journal} {Nat Commun}\ }\textbf {\bibinfo {volume} {6}},\
  (\bibinfo {year} {2015})}\BibitemShut {NoStop}%
\bibitem [{\citenamefont {Greenman}\ \emph {et~al.}(2010)\citenamefont
  {Greenman}, \citenamefont {Ho}, \citenamefont {Pabst}, \citenamefont
  {Kamarchik}, \citenamefont {Mazziotti},\ and\ \citenamefont
  {Santra}}]{GrSa-PRA-2010}%
  \BibitemOpen
  \bibfield  {author} {\bibinfo {author} {\bibfnamefont {L.}~\bibnamefont
  {Greenman}}, \bibinfo {author} {\bibfnamefont {P.~J.}\ \bibnamefont {Ho}},
  \bibinfo {author} {\bibfnamefont {S.}~\bibnamefont {Pabst}}, \bibinfo
  {author} {\bibfnamefont {E.}~\bibnamefont {Kamarchik}}, \bibinfo {author}
  {\bibfnamefont {D.~A.}\ \bibnamefont {Mazziotti}}, \ and\ \bibinfo {author}
  {\bibfnamefont {R.}~\bibnamefont {Santra}},\ }\href{\doibase
  10.1103/PhysRevA.82.023406} {\bibfield  {journal} {\bibinfo  {journal} {Phys.
  Rev. A}\ }\textbf {\bibinfo {volume} {82}},\ \bibinfo {pages} {023406}
  (\bibinfo {year} {2010})}\BibitemShut {NoStop}%
\bibitem [{\citenamefont {Pabst}(2013)}]{Pa-EPJST-2013}%
  \BibitemOpen
  \bibfield  {author} {\bibinfo {author} {\bibfnamefont {S.}~\bibnamefont
  {Pabst}},\ }\href{\doibase 10.1140/epjst/e2013-01819-x} {\bibfield  {journal}
  {\bibinfo  {journal} {Eur. Phys. J. Spec. Top.}\ }\textbf {\bibinfo {volume}
  {221}},\ \bibinfo {pages} {1} (\bibinfo {year} {2013})}\BibitemShut {NoStop}%
\bibitem [{\citenamefont {Pabst}\ \emph
  {et~al.}(2012{\natexlab{a}})\citenamefont {Pabst}, \citenamefont {Greenman},
  \citenamefont {Mazziotti},\ and\ \citenamefont {Santra}}]{PaGr-PRA-2012}%
  \BibitemOpen
  \bibfield  {author} {\bibinfo {author} {\bibfnamefont {S.}~\bibnamefont
  {Pabst}}, \bibinfo {author} {\bibfnamefont {L.}~\bibnamefont {Greenman}},
  \bibinfo {author} {\bibfnamefont {D.~A.}\ \bibnamefont {Mazziotti}}, \ and\
  \bibinfo {author} {\bibfnamefont {R.}~\bibnamefont {Santra}},\
  }\href{\doibase 10.1103/PhysRevA.85.023411} {\bibfield  {journal} {\bibinfo
  {journal} {Phys. Rev. A}\ }\textbf {\bibinfo {volume} {85}},\ \bibinfo
  {pages} {023411} (\bibinfo {year} {2012}{\natexlab{a}})}\BibitemShut
  {NoStop}%
\bibitem [{\citenamefont {Pabst}\ and\ \citenamefont
  {Santra}(2014)}]{PaSa-JPB-2014}%
  \BibitemOpen
  \bibfield  {author} {\bibinfo {author} {\bibfnamefont {S.}~\bibnamefont
  {Pabst}}\ and\ \bibinfo {author} {\bibfnamefont {R.}~\bibnamefont {Santra}},\
  }\href{\doibase 10.1088/0953-4075/47/12/124026} {\bibfield  {journal}
  {\bibinfo  {journal} {J. Phys. B}\ }\textbf {\bibinfo {volume} {47}},\
  \bibinfo {pages} {124026} (\bibinfo {year} {2014})}\BibitemShut {NoStop}%
\bibitem [{\citenamefont {Wirth}\ \emph {et~al.}(2011)\citenamefont {Wirth},
  \citenamefont {Hassan}, \citenamefont {Grguras}, \citenamefont {Gagnon},
  \citenamefont {Moulet}, \citenamefont {Luu}, \citenamefont {Pabst},
  \citenamefont {Santra}, \citenamefont {Alahmed}, \citenamefont {Azzeer},
  \citenamefont {Yakovlev}, \citenamefont {Pervak}, \citenamefont {Krausz},\
  and\ \citenamefont {Goulielmakis}}]{WiGo-Science-2011}%
  \BibitemOpen
  \bibfield  {author} {\bibinfo {author} {\bibfnamefont {A.}~\bibnamefont
  {Wirth}}, \bibinfo {author} {\bibfnamefont {M.~T.}\ \bibnamefont {Hassan}},
  \bibinfo {author} {\bibfnamefont {I.}~\bibnamefont {Grguras}},  \emph
  {et~al.},\ }\href{\doibase 10.1126/science.1210268} {\bibfield  {journal}
  {\bibinfo  {journal} {Science}\ }\textbf {\bibinfo {volume} {334}},\ \bibinfo
  {pages} {195} (\bibinfo {year} {2011})}\BibitemShut {NoStop}%
\bibitem [{\citenamefont {Pabst}\ \emph {et~al.}(2011)\citenamefont {Pabst},
  \citenamefont {Greenman}, \citenamefont {Ho}, \citenamefont {Mazziotti},\
  and\ \citenamefont {Santra}}]{PaSa-PRL-2011}%
  \BibitemOpen
  \bibfield  {author} {\bibinfo {author} {\bibfnamefont {S.}~\bibnamefont
  {Pabst}}, \bibinfo {author} {\bibfnamefont {L.}~\bibnamefont {Greenman}},
  \bibinfo {author} {\bibfnamefont {P.~J.}\ \bibnamefont {Ho}}, \bibinfo
  {author} {\bibfnamefont {D.~A.}\ \bibnamefont {Mazziotti}}, \ and\ \bibinfo
  {author} {\bibfnamefont {R.}~\bibnamefont {Santra}},\ }\href{\doibase
  10.1103/PhysRevLett.106.053003} {\bibfield  {journal} {\bibinfo  {journal}
  {Phys. Rev. Lett.}\ }\textbf {\bibinfo {volume} {106}},\ \bibinfo {pages}
  {053003} (\bibinfo {year} {2011})}\BibitemShut {NoStop}%
\bibitem [{\citenamefont {Karamatskou}\ \emph {et~al.}(2014)\citenamefont
  {Karamatskou}, \citenamefont {Pabst}, \citenamefont {Chen},\ and\
  \citenamefont {Santra}}]{KaPa-PRA-2014}%
  \BibitemOpen
  \bibfield  {author} {\bibinfo {author} {\bibfnamefont {A.}~\bibnamefont
  {Karamatskou}}, \bibinfo {author} {\bibfnamefont {S.}~\bibnamefont {Pabst}},
  \bibinfo {author} {\bibfnamefont {Y.-J.}\ \bibnamefont {Chen}}, \ and\
  \bibinfo {author} {\bibfnamefont {R.}~\bibnamefont {Santra}},\
  }\href{\doibase 10.1103/PhysRevA.89.033415} {\bibfield  {journal} {\bibinfo
  {journal} {Phys. Rev. A}\ }\textbf {\bibinfo {volume} {89}},\ \bibinfo
  {pages} {033415} (\bibinfo {year} {2014})}\BibitemShut {NoStop}%
\bibitem [{\citenamefont {Sytcheva}\ \emph {et~al.}(2012)\citenamefont
  {Sytcheva}, \citenamefont {Pabst}, \citenamefont {Son},\ and\ \citenamefont
  {Santra}}]{SyPa-PRA-2012}%
  \BibitemOpen
  \bibfield  {author} {\bibinfo {author} {\bibfnamefont {A.}~\bibnamefont
  {Sytcheva}}, \bibinfo {author} {\bibfnamefont {S.}~\bibnamefont {Pabst}},
  \bibinfo {author} {\bibfnamefont {S.-K.}\ \bibnamefont {Son}}, \ and\
  \bibinfo {author} {\bibfnamefont {R.}~\bibnamefont {Santra}},\
  }\href{\doibase 10.1103/PhysRevA.85.023414} {\bibfield  {journal} {\bibinfo
  {journal} {Phys. Rev. A}\ }\textbf {\bibinfo {volume} {85}},\ \bibinfo
  {pages} {023414} (\bibinfo {year} {2012})}\BibitemShut {NoStop}%
\bibitem [{\citenamefont {Tilley}\ \emph {et~al.}(2015)\citenamefont {Tilley},
  \citenamefont {Karamatskou},\ and\ \citenamefont {Santra}}]{TiKa-JPB-2015}%
  \BibitemOpen
  \bibfield  {author} {\bibinfo {author} {\bibfnamefont {M.}~\bibnamefont
  {Tilley}}, \bibinfo {author} {\bibfnamefont {A.}~\bibnamefont {Karamatskou}},
  \ and\ \bibinfo {author} {\bibfnamefont {R.}~\bibnamefont {Santra}},\
  }\href{http://stacks.iop.org/0953-4075/48/i=12/a=124001} {\bibfield
  {journal} {\bibinfo  {journal} {Journal of Physics B: Atomic, Molecular and
  Optical Physics}\ }\textbf {\bibinfo {volume} {48}},\ \bibinfo {pages}
  {124001} (\bibinfo {year} {2015})}\BibitemShut {NoStop}%
\bibitem [{\citenamefont {Starace}(1970)}]{St-PRA-1970}%
  \BibitemOpen
  \bibfield  {author} {\bibinfo {author} {\bibfnamefont {A.~F.}\ \bibnamefont
  {Starace}},\ }\href{\doibase 10.1103/PhysRevA.2.118} {\bibfield  {journal}
  {\bibinfo  {journal} {Phys. Rev. A}\ }\textbf {\bibinfo {volume} {2}},\
  \bibinfo {pages} {118} (\bibinfo {year} {1970})}\BibitemShut {NoStop}%
\bibitem [{Note1()}]{Note1}%
  \BibitemOpen
  \bibinfo {note} {S. Pabst, L. Greenman, A. Karamatskou, Y.-J. Chen, A.
  Sytcheva, O. Geffert, R. Santra--\protect \textsc {xcid} program package for
  multichannel ionization dynamics, DESY, Hamburg, Germany, 2015, Rev.
  1790}\BibitemShut {NoStop}%
\bibitem [{Note2()}]{Note2}%
  \BibitemOpen
  \bibinfo {note} {A pseudo-spectral grid with a radial box size of 100~$a_0$,
  500 grid points, and a mapping parameter of $\zeta =0.5$ are used. The
  complex absorbing potential starts at 70~$a_0$ and has a strength of $\eta
  =0.002$. The maximum angular momentum is 6 and Hartree-Fock orbitals up to an
  energy of 200~$E_h$ are considered. The propagation method is Runge-Kutta 4
  with a time step $dt=0.005$~a.u. All $4d, 5s$, and $5p$ orbitals are active
  in the calculations.}\BibitemShut {Stop}%
\bibitem [{\citenamefont {Pabst}\ \emph
  {et~al.}(2012{\natexlab{b}})\citenamefont {Pabst}, \citenamefont {Sytcheva},
  \citenamefont {Moulet}, \citenamefont {Wirth}, \citenamefont {Goulielmakis},\
  and\ \citenamefont {Santra}}]{PaSy-PRA-2012}%
  \BibitemOpen
  \bibfield  {author} {\bibinfo {author} {\bibfnamefont {S.}~\bibnamefont
  {Pabst}}, \bibinfo {author} {\bibfnamefont {A.}~\bibnamefont {Sytcheva}},
  \bibinfo {author} {\bibfnamefont {A.}~\bibnamefont {Moulet}}, \bibinfo
  {author} {\bibfnamefont {A.}~\bibnamefont {Wirth}}, \bibinfo {author}
  {\bibfnamefont {E.}~\bibnamefont {Goulielmakis}}, \ and\ \bibinfo {author}
  {\bibfnamefont {R.}~\bibnamefont {Santra}},\ }\href{\doibase
  10.1103/PhysRevA.86.063411} {\bibfield  {journal} {\bibinfo  {journal} {Phys.
  Rev. A}\ }\textbf {\bibinfo {volume} {86}},\ \bibinfo {pages} {063411}
  (\bibinfo {year} {2012}{\natexlab{b}})}\BibitemShut {NoStop}%
\bibitem [{\citenamefont {Delone}\ and\ \citenamefont
  {Krainov}(2000)}]{DeKr-book}%
  \BibitemOpen
  \bibfield  {author} {\bibinfo {author} {\bibfnamefont {N.~B.}\ \bibnamefont
  {Delone}}\ and\ \bibinfo {author} {\bibfnamefont {V.~P.}\ \bibnamefont
  {Krainov}},\ }\href@noop {} {\emph {\bibinfo {title} {Multiphoton Processes
  in Atoms}}}\ (\bibinfo  {publisher} {Springer, Berlin},\ \bibinfo {year}
  {2000})\BibitemShut {NoStop}%
\bibitem [{\citenamefont {Freeman}\ \emph {et~al.}(1987)\citenamefont
  {Freeman}, \citenamefont {Bucksbaum}, \citenamefont {Milchberg},
  \citenamefont {Darack}, \citenamefont {Schumacher},\ and\ \citenamefont
  {Geusic}}]{FrBu-PRL-1987}%
  \BibitemOpen
  \bibfield  {author} {\bibinfo {author} {\bibfnamefont {R.~R.}\ \bibnamefont
  {Freeman}}, \bibinfo {author} {\bibfnamefont {P.~H.}\ \bibnamefont
  {Bucksbaum}}, \bibinfo {author} {\bibfnamefont {H.}~\bibnamefont
  {Milchberg}}, \bibinfo {author} {\bibfnamefont {S.}~\bibnamefont {Darack}},
  \bibinfo {author} {\bibfnamefont {D.}~\bibnamefont {Schumacher}}, \ and\
  \bibinfo {author} {\bibfnamefont {M.~E.}\ \bibnamefont {Geusic}},\
  }\href{\doibase 10.1103/PhysRevLett.59.1092} {\bibfield  {journal} {\bibinfo
  {journal} {Phys. Rev. Lett.}\ }\textbf {\bibinfo {volume} {59}},\ \bibinfo
  {pages} {1092} (\bibinfo {year} {1987})}\BibitemShut {NoStop}%
\bibitem [{\citenamefont {Fleischhauer}\ \emph {et~al.}(2005)\citenamefont
  {Fleischhauer}, \citenamefont {Imamoglu},\ and\ \citenamefont
  {Marangos}}]{FlIm-RMP-2005}%
  \BibitemOpen
  \bibfield  {author} {\bibinfo {author} {\bibfnamefont {M.}~\bibnamefont
  {Fleischhauer}}, \bibinfo {author} {\bibfnamefont {A.}~\bibnamefont
  {Imamoglu}}, \ and\ \bibinfo {author} {\bibfnamefont {J.~P.}\ \bibnamefont
  {Marangos}},\ }\href{\doibase 10.1103/RevModPhys.77.633} {\bibfield
  {journal} {\bibinfo  {journal} {Rev. Mod. Phys.}\ }\textbf {\bibinfo {volume}
  {77}},\ \bibinfo {pages} {633} (\bibinfo {year} {2005})}\BibitemShut
  {NoStop}%
\bibitem [{\citenamefont {Rza\ifmmode~\mbox{\c{}}\else \c{}\fi{}ewski}\ \emph
  {et~al.}(1985)\citenamefont {Rza\ifmmode~\mbox{\c{}}\else \c{}\fi{}ewski},
  \citenamefont {Zakrzewski}, \citenamefont {Lewenstein},\ and\ \citenamefont
  {Haus}}]{RzZa-PRA-1985}%
  \BibitemOpen
  \bibfield  {author} {\bibinfo {author} {\bibfnamefont {K.}~\bibnamefont
  {Rza\ifmmode~\mbox{\c{}}\else \c{}\fi{}ewski}}, \bibinfo {author}
  {\bibfnamefont {J.}~\bibnamefont {Zakrzewski}}, \bibinfo {author}
  {\bibfnamefont {M.}~\bibnamefont {Lewenstein}}, \ and\ \bibinfo {author}
  {\bibfnamefont {J.~W.}\ \bibnamefont {Haus}},\ }\href{\doibase
  10.1103/PhysRevA.31.2995} {\bibfield  {journal} {\bibinfo  {journal} {Phys.
  Rev. A}\ }\textbf {\bibinfo {volume} {31}},\ \bibinfo {pages} {2995}
  (\bibinfo {year} {1985})}\BibitemShut {NoStop}%
\end{thebibliography}%


\end{document}